\input{aipcheck}

\documentclass[
    ,final            % use final for the camera ready runs
  ]
  {aipproc}

\layoutstyle{6x9}

\usepackage{rotating}

\begin{document}

\def\met{\mbox{\ensuremath{\, \slash\kern-.6emE_{\rm T}}}}

\title{Fake $\met{}$ from Calorimeter Effects}


\author{Bernhard Meirose \\ 
On behalf of the ATLAS Collaboration
}{
  address={Albert-Ludwigs-Universit\"{a}t Freiburg \\
Hermann-Herder-Str. 3 79104 Freiburg i. Br., Germany}
}

\begin{abstract}

In this paper we discuss briefly the correlation between fake $\met{}$ and jets pointing to crack regions in the ATLAS calorimeters.

\end{abstract}

\maketitle

%%%%%%%%%%%%%%%%%%%%%%%%%%%%%%%%%%%%%%%%%%%%
%% MAINMATTER
%%%%%%%%%%%%%%%%%%%%%%%%%%%%%%%%%%%%%%%%%%%%

\section{Introduction}

One of the key signatures for new physics, such as supersymmetry (SUSY), in the ATLAS detector \cite{ATLAS}, rely on a good measurement of the missing transverse energy ($\met{}$). Therefore any sources of fake $\met{}$ generate backgrounds to SUSY. Examples are: beam-gas interactions, beam halo muons, cosmic rays and electronics noise. Here we discuss fake $\met{}$  due to calorimeter imperfections \cite{CSC}.

\section{Fake $\met{}$ from calorimeter transition regions}

In what follows we assume all calorimeter readout channels are functional. Fake $\met{}$ in the calorimeter
is then produced by mis-measurements of hadronic jets, taus, electrons or photons, mostly caused by un-instrumented detector regions. In ATLAS these pseudorapidity ($\eta$) regions are: (1.3 $<$ |$\eta$| $<$ 1.6) and (3.1 $<$ |$\eta$|$<$ 3.3). Figure 1 shows the $\eta$ distribution of the worst and the second worst measured jet (defined w.r.t. the closest true jet and their energy difference) for QCD dijets generated with 560 $<$ $p_T$ $<$ 1120 GeV. It shows that a large number of the worst measured jets have $\eta$ in 1.3-1.6. The distribution of the second worst measured jet peaks around |$\eta$| in 0.6-0.9.

\begin{figure}[!h]
  \includegraphics[height=.25\textheight]{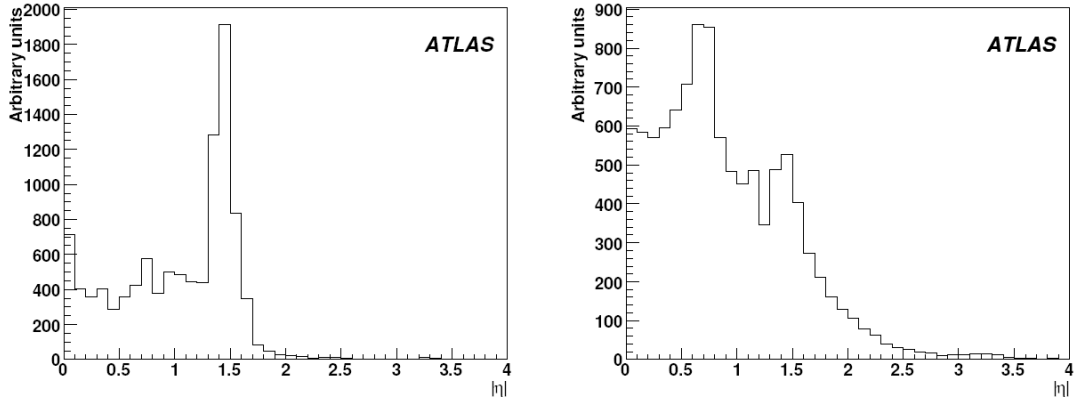}
  \caption{The $\eta$ distribution of the worst (left) and second-worst measured jet (right) in the calorimeter
in QCD events generated with 560 < $p_T$ < 1120 GeV and with a fake $\met{}$ larger than 100 GeV.}
\end{figure}

The above result suggests a large correlation between the jet $\eta$ and fake $\met{}$. However, Fig. 2 shows the fake $\met{}$ distribution for QCD jets generated with 560 < $p_T$ < 1120 GeV and 140 < $p_T$ < 280 GeV, when a jet points to the crack/gap region or not. The slope of the distribution suggests no significant correlation between jets pointing to cracks and fake $\met{}$.

This apparent contradiction can be understood as follows: even though the worst measured jet contributes strongly to the fake $\met{}$, it is not the only source of fake $\met{}$ in the event. The worst and the second worst measured jets contribute on average about 60$\%$ and 20$\%$ to fake $\met{}$, respectively. But not all worst measured jets are along the crack region and each event has many jets. In lower fake $\met{}$ regions there is a stronger correlation between fake $\met{}$ and jets pointing to cracks. For higher fake $\met{}$ there is more than one source contributing and the correlation of jets pointing to cracks is smeared out as can be seen in Fig. 2.

In conclusion, although together the worst measured jets contribute on average about 80$\%$ to fake $\met{}$ in the event, no significant correlation between jets pointing to cracks and fake $\met{}$ is found. The reason for this is that not all worst measured jets are along the crack region and each event has many jets. Thererefore in regions of large jet multiplicity the correlation between fake $\met{}$ and jets pointing to cracks cannot be seen.

\begin{figure}[!h]
  \includegraphics[height=.25\textheight]{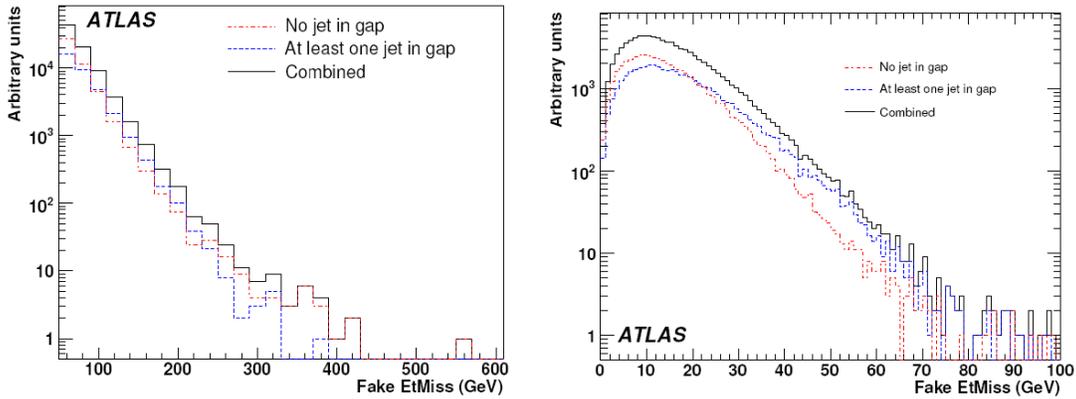}
  \caption{The fake $\met{}$ rate for QCD sample in 560 < $p_T$ < 1120 GeV range (left) and QCD sample in 140 < $p_T$  < 280 GeV (right) due to calorimeter mis-measurements.}
\end{figure}

%\section{Conclusion}

%Although together the worst measured jets contribute on average about 80$\%$ to fake $\met{}$ in the event, no significant correlation between jets pointing to cracks and fake $\met{}$ is found. The reason for this apparent contradiction is that not all worst measured jets are along the crack region and each event has many
%jets. Thererefore in regions of large jet multiplicity the correlation between fake $\met{}$ and jets pointing to cracks is smeared out.

%%%%%%%%%%%%%%%%%%%%%%%%%%%%%%%%%%%%%%%%%%%%%%%%
%% BACKMATTER
%%%%%%%%%%%%%%%%%%%%%%%%%%%%%%%%%%%%%%%%%%%%%%%%

%\begin{theacknowledgments}
%This work was supported by the Federal Ministry of Education and Research (Germany).
%\end{theacknowledgments}


\begin{thebibliography}{9}

\bibitem{ATLAS}
G. Aad et al. (The ATLAS Collaboration), The ATLAS Experiment at the CERN Large Hadron Collider, 2008. 437 pp. Published in JINST 3:S08003,2008.


\bibitem{CSC}
G. Aad et al. (The ATLAS Collaboration), Expected Performance of the ATLAS Experiment - Detector, Trigger and Physics,
Jan 2009. 1852pp. 

\end{thebibliography}
\end{document}